\documentclass[aps,pra,twocolumn,a4paper,groupedaddress,floatfix,10pt,longbibliography]{revtex4-1}
\usepackage{graphicx,amsmath,amssymb,amsfonts,dsfont,enumitem,MnSymbol}
\usepackage[utf8]{inputenc}
\newcommand{\mf}[1]{\boldsymbol{#1}}
\newcommand{\ket}[1]{\ensuremath{|#1\rangle}}
\newcommand{\mc}[1]{\ensuremath{\mathcal{#1}}}
\newcommand{\bra}[1]{\ensuremath{\langle #1 |}}

\newcommand{\imag}{\mathrm{i}}
\usepackage{color}

\setitemize{leftmargin=*} 

\begin{document} 

\title{
Manipulating quantum materials with quantum light
}
\author{Martin Kiffner${}^{1,2}$}
\author{Jonathan R. Coulthard${}^{2}$}
\author{Frank Schlawin${}^{2}$}
\author{Arzhang Ardavan${}^{2}$}
\author{Dieter Jaksch${}^{2,1}$}

\affiliation{Centre for Quantum Technologies, National University of Singapore,
3 Science Drive 2, Singapore 117543${}^1$}
\affiliation{Clarendon Laboratory, University of Oxford, Parks Road, Oxford OX1
3PU, United Kingdom${}^2$}

\begin{abstract}
We show that the macroscopic magnetic and electronic properties of strongly correlated electron systems 
can be manipulated by  coupling them to a  cavity mode. 
As a paradigmatic example we consider the Fermi-Hubbard model 
and  find that the electron-cavity coupling 
enhances the magnetic interaction between the electron spins in the ground-state manifold. At half filling this effect can be observed by a change in the magnetic 
susceptibility. At less than half filling, the cavity 
introduces a next-nearest neighbour hopping and mediates a long-range electron-electron interaction between distant sites. 
We study the ground state properties 
with Tensor Network methods and find that the cavity coupling 
can induce a new phase characterized by a momentum-space pairing effect for electrons.
\end{abstract}

\maketitle

\section{Introduction \label{introduction}}
The ability to  control and manipulate complex quantum systems  is of paramount importance for 
future quantum technologies. Of particular interest are 
quantum hybrid systems~\cite{xiang:13,kurizki:15} where  different quantum objects 
hybridize to exhibit properties not shared by the individual components. 
Examples of this hybridization effect in cold atom systems 
coupled to optical cavities  are given by  self-organization 
phenomena~\cite{domokos:02,nagy:08,black:03,ritsch:13,piazza:13} as well as the occurrence  of quantum phase transitions 
and exotic quantum phases~\cite{jaksch:01,baumann:10,landig:16,kollath:16}.
Recently, the class of available hybrid systems has been extended   by solid state systems that couple strongly to  microwave, terahertz or optical 
radiation~\cite{zhang:14,tabuchi:14,yao:15,sivarajah:17b,abdurakhimov:18,mergenthaler:17,hagenmueller:10,scalari:12,zhang:16,li:18,paravicini:18,bartolo:18,orgiu:15,feist:15,schachenmayer:15,laplace:16,schlawin:18,curtis:18,mazza:18}.  
For example, the coupling of microwave cavities to magnon and spinon excitations in magnetic materials has been investigated in~\cite{zhang:14,tabuchi:14,yao:15,sivarajah:17b,abdurakhimov:18} 
and~\cite{mergenthaler:17}, respectively. Two-dimensional electron gases in magnetic fields can couple very strongly to terahertz cavities~\cite{hagenmueller:10,scalari:12,zhang:16} 
such that Bloch-Siegert shifts become observable~\cite{li:18}, and tomography of an ultrastrongly coupled polariton state was presented in~\cite{paravicini:18} 
using magneto-transport measurements~\cite{bartolo:18}. 
A recent experiment~\cite{orgiu:15} has demonstrated  that coupling of an organic semiconductor to an optical cavity enhances the electric conductivity, 
which can be understood in terms of delocalized exciton polaritons~\cite{feist:15,schachenmayer:15}. 
A special class of solid state systems are quantum materials~\cite{editorial:16,powell:06,powell:11,kato:04} where 
small microscopic changes can result in large macroscopic responses due to strong electron-electron interactions. Coupling these systems to cavities 
opens up the fascinating possibility of  investigating the ultimate quantum limit where macroscopic properties of quantum materials are 
determined by quantum light fields and vice versa. 
First steps into this direction have been undertaken recently~\cite{laplace:16,schlawin:18,curtis:18,sentef:18,mazza:18}. For example, quantum counterparts of light-induced 
superconductivity ~\cite{fausti:11,hu:14,mitrano:16,sentef:17} have been investigated 
in~\cite{laplace:16,schlawin:18,curtis:18,sentef:18} using terahertz and microwave cavities, and a superradiant phase of a cavity-coupled quantum material 
has been predicted in~\cite{mazza:18}. 

Here we show that shaping the vacuum via a cavity allows one to manipulate macroscopic properties like the magnetic susceptibility 
 of a quantum material. As a paradigmatic model of quantum materials we consider the Fermi-Hubbard model~\cite{essler:05} which captures the 
interplay between kinetic fluctuations and strong, local, electron-electron interactions~\cite{powell:06,powell:11,kato:04,kato:04}.  
While the interaction of these systems with strong, classical light fields has been investigated, for example, in~\cite{mentink:15,coulthard:17,goerg:18,stepanov:17}, 
the intriguing possibility of coupling them to quantum light fields has not been explored yet. 
%
%%%%%%%%%%%%%%%%%%%%%%
\begin{figure}[t!]
\begin{center}
\includegraphics[width=\columnwidth]{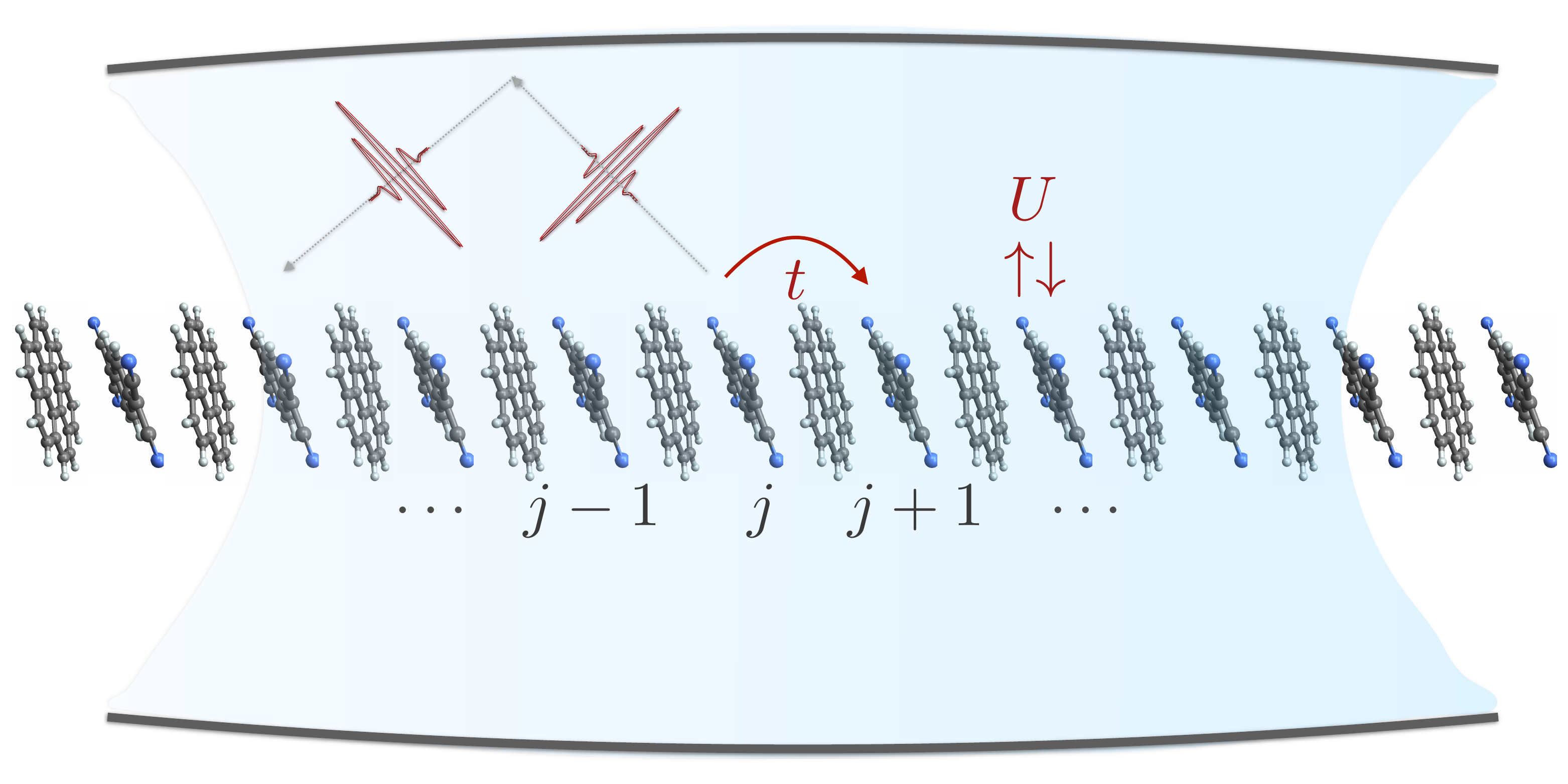}
\end{center}
\caption{\label{fig1}
(Color online) 
The system of interest is given by an electronic system coupled to a single-mode cavity. The electronic system 
is described by the Fermi-Hubbard model with on-site interaction $U$ and hopping amplitude $t$ between 
neighbouring sites. 
}
\end{figure}
%%%%%%%%%%%%%%%%%%%%
% 

More specifically, we consider a one-dimensional Hubbard model coupled to a single mode of an empty cavity. 
For an electronic system at half filling we find that  the electron-cavity interaction  enhances the magnetic interactions between 
spins in the ground-state manifold. This effect can be experimentally observed by measuring the magnetic susceptibility. 

At less than half filling, the cavity coupling  introduces (i) a next-nearest neighbour hopping, (ii) an on-site energy shift  and (iii) a long-range electron-electron interaction between distant sites. 
We investigate the ground state of the electronic system at less than half filling and in the presence of the cavity with density matrix renormalization group (DMRG) techniques~\cite{schollwock:11,TNT}. We find that the cavity induces  momentum-space pairing for mesoscopic electron systems. 
The transition to this new phase is a collectively enhanced effect and does not require ultra-strong coupling on the single-electron level and scales
with $1/\sqrt{v_{\text{uc}}}$, where $v_{\text{uc}}$ is the volume of the unit cell of the crystal.

This paper is organized as follows. In Sec.~\ref{model} we introduce our model 
for the system shown in Fig.~\ref{fig1}. Our results are presented in 
Sec.~\ref{results}, and their experimental realization is discussed in 
Sec.~\ref{exp}. A brief discussion and conclusion is provided in 
Sec.~\ref{conclusion}. 
\section{Model \label{model}}
The system shown in Fig.~\ref{fig1} is comprised of an electronic system coupled to a single-mode cavity. 
We introduce the Hamiltonian describing this quantum hybrid system in Sec.~\ref{sysham}, and discuss 
its gross energy structure for the parameters of interest  in Sec.~\ref{perturb}. 
\subsection{System Hamiltonian \label{sysham}}
We begin with the description of the system Hamiltonian with the single-mode cavity with resonance frequency $\omega_c$. 
The Hamiltonian for the cavity photons with energy  $\Omega =\hbar\omega_c$ is 
\begin{align}
 \hat{P} = \Omega \hat{a}^{\dagger} \hat{a},
\label{hp}
\end{align}
where $\hat{a}^{\dagger}$ ($\hat{a}$) is the bosonic photon creation (annihilation) operator. 
The eigenstates of $\hat{P}$ are the photon number states $\ket{j_{P}}$ with $\hat{P}\ket{j_{P}} = j \Omega \ket{j_{P}}$. 
The spectral decomposition of $\hat{P}$ can thus be written as
\begin{align}
 \hat{P} = \Omega \sum\limits_{j = 0}^{\infty} j \hat{\mc{P}}_j^{P}\,,
\end{align}
where
\begin{align}
 \hat{\mc{P}}_j^{P} = \ket{j_{P}}\bra{j_{P}}
 \label{pjP}
\end{align}
is the projector onto the state  with $j$ photons.  
The electronic system is described by the one-dimensional Fermi-Hubbard model~\cite{essler:05} 
with Hamiltonian 
\begin{align}
 \hat{H}_{\text{FH}} = \hat{T}  +  \hat{D}   \,,
 \label{HFH}
\end{align}
where
\begin{align}
 \hat{T} = & -t \sum\limits_{ \langle jk\rangle \sigma} \left(\hat{c}_{j,\sigma}^{\dagger}\hat{c}_{k,\sigma} 
 + \text{h.c.}\right)  
 \label{T}
 \end{align}
  accounts for hopping between neighbouring sites $\langle j k\rangle$ with $j<k$, $t$ is  the hopping amplitude 
 and $\hat{c}_{j,\sigma}^{\dagger}$ ($\hat{c}_{j,\sigma}$) creates (annihilates) an electron at 
 site $j$ in spin state $\sigma\in\{\uparrow,\downarrow\}$. 
 The second term in Eq.~(\ref{HFH}) describes the on-site Coulomb interaction between electrons, 
 \begin{align}
 \hat{D} =& U \sum\limits_{j} \hat{n}_{j,\uparrow}\hat{n}_{j,\downarrow}\,,
 \label{D}
 \end{align}
where $U$ is the  interaction energy and 
$\hat{n}_{j,\sigma}=\hat{c}_{j,\sigma}^{\dagger}\hat{c}_{j,\sigma}$ counts the number of electrons at 
site $j$ in spin state $\sigma$. 
Each site can accommodate at most two electrons with opposite spins, and 
in the following we refer to   doubly occupied sites as  doublons. 

The operator $\hat{D}$ is diagonal in the basis of Wannier states~\cite{essler:05}, 
\begin{align}
\ket{\mf{x},\mf{s}} = \hat{c}_{x_N,s_N}^{\dagger}\ldots\hat{c}_{x_1,s_1}^{\dagger}\ket{0_E}\,,
\label{wannier}
\end{align}
where $\ket{0_E}$ is the vacuum state of the electronic system and 
\begin{subequations}
 \label{dist}
\begin{align}
 \mf{x} & = (x_1,\ldots,x_N)\,,\\
 \mf{s} & = (s_1,\ldots,s_N)\,,
\end{align}
\end{subequations}
are row vectors with $x_j \in\{1,\ldots,L\}$, $s_j \in\{\uparrow,\downarrow\}$ and  $j \in\{1,\ldots,N\}$. 
The vectors in Eq.~(\ref{dist}) describe the spatial distribution of  $N$ electrons  and their spin state 
in a one-dimensional lattice with $L$ sites. 

The Wannier states are eigenstates of $\hat{D}$ and form 
degenerate manifolds with energies $k U$, where $k$ is an integer 
that counts the total number of doubly occupied sites in $\ket{\mf{x},\mf{s}}$. 
In the following we refer to   doubly occupied sites as  doublons. 
The projector  
onto the manifold with $k$ doublons is given by~\cite{essler:05}
\begin{align}
 \hat{\mc{P}}_k^{D} = \left.\frac{(-1)^k}{k!}\partial_{x}^k G(x)\right|_{x=1}\, ,
 \label{pjD}
\end{align}
where 
\begin{align}
 G(x) = \prod\limits_{j=1}^L  \left( 1- x \,\hat{n}_{j,\uparrow}\hat{n}_{j,\downarrow} \right)
\end{align}
is the generating function. The spectral decomposition of $\hat{D}$ is thus given by 
 \begin{align}
 \hat{D} =& U \sum\limits_{k=0}^{L} k\, \hat{\mc{P}}_k^{D} \,.
 \label{Ds}
 \end{align}
Note that in general, each manifold with a given number of doublons contains a large number of electronic states. 
For example,  the  ground state manifold  with no doublons contains 
 \begin{align}
  \#\left(\hat{\mc{P}}_0^{D}\right)  = 2^N\left( \genfrac{}{}{0pt}{}{L}{N}\right)
 \end{align}
states for a system with $N\le L$ electrons. 
The preceding definitions for $\hat{H}_{\text{FH}}$ and $ \hat{P}$ allow  us to 
write the total Hamiltonian of the hybrid system  in Fig.~\ref{fig1} as 
\begin{align}
 \hat{H} & = \hat{H}_{\text{FH}} + \hat{P}   + \hat{V}\,, % \\
  %
 % & = \hat{H}_0 + \hat{H}_1\,, 
 %
 \label{totv}
\end{align}
where $\hat{V}$ accounts for the electron-photon interaction. 
In Appendix~\ref{elphot} we outline 
the derivation of this interaction term from first principles and find 
\begin{align}
 \hat{V}=  g (\hat{a}+\hat{a}^{\dagger}) \hat{\mc{J}}\,,
\label{Vint}
\end{align}
where
 \begin{align}
 \hat{\mc{J}} = &-\imag  \sum\limits_{\langle j k \rangle \sigma } \left(\hat{c}_{j,\sigma}^{\dagger}\hat{c}_{k,\sigma} 
 -\hat{c}_{k,\sigma}^{\dagger}\hat{c}_{j,\sigma} \right) 
 \label{current}
\end{align}
is the dimensionless current operator. 
The parameter $g=t \eta$ in $\hat{V}$ determines the coupling strength between the electrons and photons, 
and  the dimensionless parameter 
\begin{align}
\eta=\frac{d e}{\sqrt{2\hbar\varepsilon_0 \omega_c v}} 
\label{eta}
\end{align}
depends on the lattice constant $d$ and  the cavity  mode volume $v$ 
($e$: elementary charge, $\varepsilon_0$: vacuum permittivity, $\hbar$: reduced Planck's constant). 
The cavity mode couples to both spin components of the electrons in the same way,  and the derivation of $\hat{V}$  assumes~$\eta \ll 1$. 
Furthermore, we point out that $\hat{V}$ in Eq.~(\ref{Vint}) is fundamentally different 
from cold atom systems where the light-matter 
coupling is proportional to the atomic density rather than the current. 
%
% 
%%%%%%%%%%%%%%%%%%%%%%
\begin{figure}[t!]
\begin{center}
\includegraphics[width=\columnwidth]{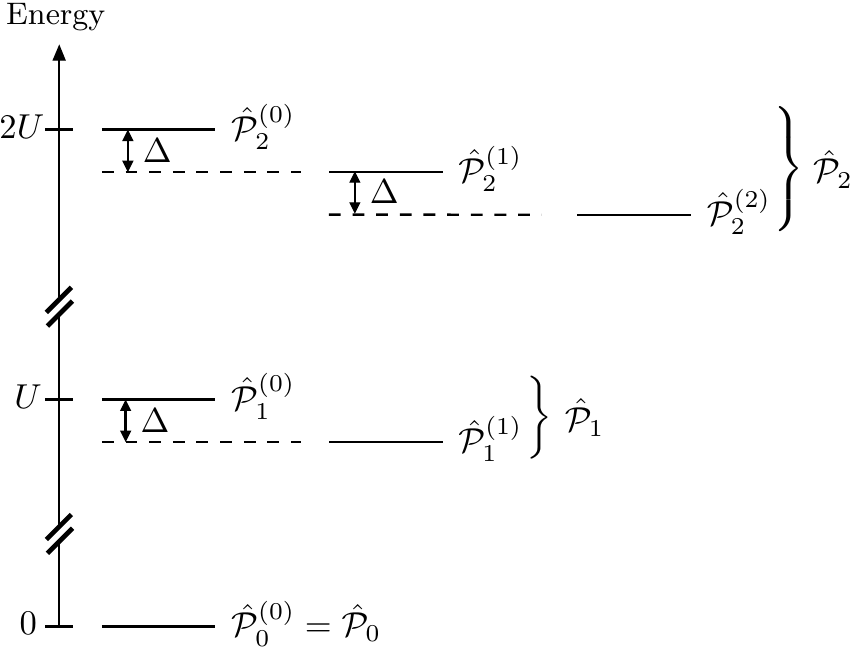}
\end{center}
\caption{\label{fig2}
 Schematic drawing of the spectrum of $\hat{H}_0 = \hat{D}  + \hat{P}$. $\Delta = \Omega - U$ is 
the difference between the photon and doublon energies,  $\hat{\mc{P}}_n^{(j)}$ projects onto  
a sub-manifold with $j$ photons and $n-j$ doublons, 
and $\hat{\mc{P}}_n =  \sum_{j=0}^{n} \hat{\mc{P}}_n^{(j)}$.  
Higher excitations not shown.
}
\end{figure}
%%%%%%%%%%%%%%%%%%%%
% 
%
\subsection{Gross energy structure of $\hat{H}$ \label{perturb}}
Throughout this work we assume that the photon energy $\Omega$ is of the 
same order of magnitude as the interaction energy $U$ of doubly occupied sites. Since we assume a strongly correlated electron system with 
$U\gg t$ and since the electron-photon coupling obeys $t \gg g$, we have $U, \,\Omega \gg t, g$. 
This separation of energy scales suggests writing the Hamiltonian in Eq.~(\ref{totv}) as $\hat{H}= \hat{H}_0 +\hat{H}_1$,  where 
 \begin{align}
 \hat{H}_0 & = \hat{D} + \hat{P}  
 \label{H0}
 \end{align}
 describes the energy of doublons and photons,  and 
 \begin{align}
  \hat{H}_1 & = \hat{T} + \hat{V}  
 \label{H1}
\end{align}
accounts for the kinetic energy and the electron-photon interaction. 

Next we investigate the spectrum of $\hat{H}_0$ in more detail. Due to the structure of $\hat{H}_0$ its eigenstates 
$\ket{\mf{x},\mf{s}}\otimes\ket{j_P}$ are the tensor product of eigenstates of $\hat{D}$ and $\hat{P}$. 
The energies of the states $\ket{\mf{x},\mf{s}}$ and $\ket{j_P}$ are determined by their  number of doublons and photons, respectively. 
Here we group the eigenstates of $\hat{H}_0$ into manifolds with  the same  number of excitations  as shown in 
Fig.~\ref{fig2}, where an excitation can be either a photon or a doublon. 
There are $n+1$ possibilities to form $n$ excitations out of doublons and  photons, 
and these decompositions have  energies 
\begin{align}
 E_n^{(j)} = (n-j)U + j \Omega =n U + j \Delta \, ,
 \label{enj}
\end{align}
where  $\Delta = \Omega -U$ is the energy difference between the cavity and the doublon transition, $j\in\{0,\ldots,n\}$ denotes the number of photons and $n-j$ is the number of doublons. 
The corresponding projector onto the sub-manifold with energy $E_n^{(j)}$ is 
\begin{align}
 \hat{\mc{P}}_n^{(j)} = \hat{\mc{P}}_{n -j }^{D}\otimes \hat{\mc{P}}_{j}^{P} \,,
 \label{pnj}
\end{align}
where $ \hat{\mc{P}}_k^{D}$ and $\hat{\mc{P}}_j^{P}$ are defined in    Eqs.~(\ref{pjD}) and~(\ref{pjP}), respectively.
Finally, we introduce 
\begin{align}
 \hat{\mc{P}}_n =  \sum\limits_{i=0}^{n} \hat{\mc{P}}_n^{(i)} \,,
 \label{pn}
\end{align}
which is  the projector  onto the manifold with $n$ excitations. 
\section{Results \label{results}}
In Sec.~\ref{model} we have shown that the Hamiltonian describing the system in Fig.~\ref{fig1} 
can be split in two terms that correspond to different energy scales of the problem. The first, large  
energy scale is given by the electron-electron interaction and the cavity frequency. The second, small 
energy scale is the hopping amplitude and the electron-cavity coupling. 
Due to this separation of energy scales we can investigate the physics of the low-energy sector of the system in an 
effective Hamiltonian approach described in Sec.~\ref{effham}. Results of a numerical investigation of the ground state of this effective 
Hamiltonian are presented in Sec.~\ref{numerics}. 
\subsection{Effective Hamiltonian \label{effham}}
Here we investigate the physics of the electron-photon coupling in the low-energy manifold $\mc{P}_0$ with zero  excitations. 
The effective  Hamiltonian in this ground-state manifold and in second-order perturbation theory is given by~\cite{tannoudji:api}
 \begin{align}
  \hat{H}_{\text{gs}} = \hat{\mc{P}}_0 \hat{H}_1 \hat{\mc{P}}_0 + 
  \sum\limits_{\begin{subarray}{c} m\\ m\not=0 \end{subarray}} 
  \sum\limits_{j=0}^m
  \frac{\hat{\mc{P}}_0 \hat{H}_1 \hat{\mc{P}}_m^{(j)} \hat{H}_1 \hat{\mc{P}}_0}{E_0-E_m^{(j)}} \,, 
  \label{gs}
 \end{align}
where $\hat{H}_1$, $E_m^{(j)}$ and  $\hat{\mc{P}}_m^{(j)}$ are defined in Eqs.~(\ref{H1}),~(\ref{enj}) and~(\ref{pnj}), respectively. 
 $E_0$  is the energy of states  in the $\hat{\mc{P}}_0$ manifold with respect to $\hat{H}_0$, and we set $E_0=0$ in the following.

\begin{figure}[t!]
\begin{center}
\includegraphics[width=\columnwidth]{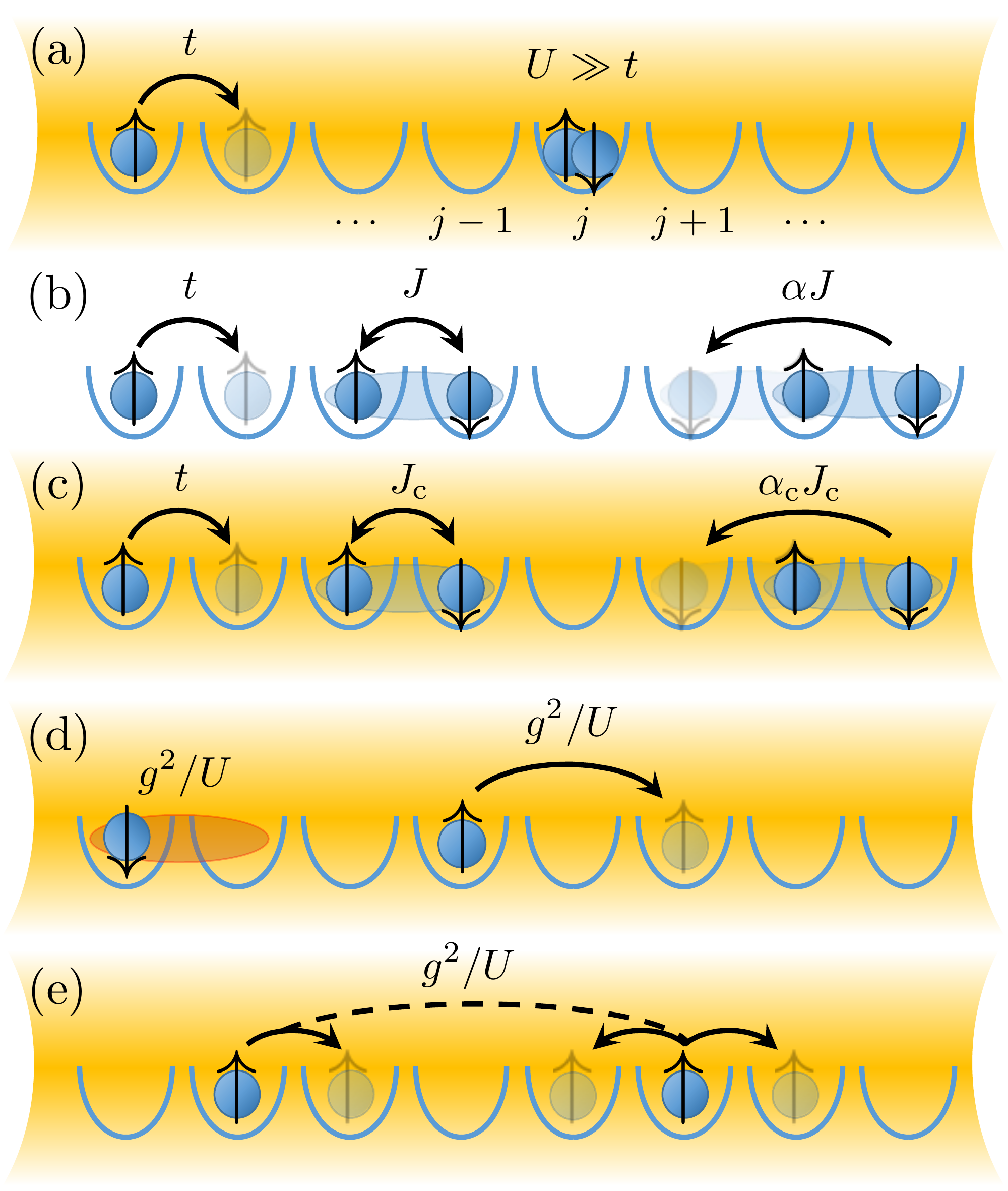}
\end{center}
\caption{\label{fig3}
(Color online) (a) Electronic system coupled to a single-mode cavity. The electronic system 
 is described by a one-dimensional Fermi-Hubbard model with on-site interaction $U$ and hopping amplitude $t$. 
(b) Processes in the $tJ\alpha$ model 
without the cavity where $J$ is the magnetic interaction and $\alpha J$ is the pair hopping amplitude. 
(c) The cavity  leaves the hopping amplitude $t$ unchanged but gives rise to re-normalized  parameters $J_c$ and $\alpha_c$. 
(d) The cavity introduces a particle-hole binding  and enables a next-nearest neighbour hopping term, and 
(e) enables a long-range electron-electron interaction.
}
\end{figure}

 In order to clearly single out the effect of the cavity on the electronic system, we begin with a discussion of the 
 case where the electron-photon coupling is zero, i.e.,   $g=0$. In this case only the $m=0$, $j=0$ term  contributes in the sum in 
 Eq.~(\ref{gs}), and $ \hat{H}_{\text{gs}}$ reduces to the well-known $tJ\alpha$ model~\cite{essler:05}, 
\begin{align}
  \hat{H}[tJ\alpha] = \hat{\mc{P}}_0\left(\hat{T} + \hat{H}_{\text{mag}}[J]  + \hat{H}_{\text{pair}}[\alpha J]   \right) \hat{\mc{P}}_0\,, 
 \label{res0}
\end{align}
where  
\begin{subequations}
\begin{align}
\hat{H}_{\text{mag}}[J] = & -J \sum\limits_{\langle kl \rangle} \hat{b}_{kl}^{\dagger}\hat{b}_{kl} \,, \label{Hex} \\
\hat{H}_{\text{pair}}[\alpha J] = &
-\alpha J \sum\limits_{\llangle klj\rrangle} \left( \hat{b}_{kl}^{\dagger}\hat{b}_{lj} +\text{h.c.}\right) \,,
\label{Hpair}
\end{align}
\end{subequations}
 and 
 \begin{align}
 \hat{b}_{kl}^{\dagger} = \left( \hat{c}_{k,\uparrow}^{\dagger} \hat{c}_{l,\downarrow}^{\dagger} 
 - \hat{c}_{k,\downarrow}^{\dagger} \hat{c}_{l,\uparrow}^{\dagger}\right)/\sqrt{2} 
\label{bkl}
\end{align}
creates a singlet pair at sites $k$ and $l$. 
The physical processes of the $tJ\alpha$ model are illustrated in  Fig.~\ref{fig3}(b). $\hat{T}$ describes the  
hopping between adjacent sites, and  $\hat{H}_{\text{mag}}$ binds nearest-neighbour singlet pairs with 
energy $J=4  t^2/U$ via superexchange processes. 
$\hat{H}_{\text{pair}}$  describes the hopping of   singlet pairs on neighbouring sites as shown in Fig.~\ref{fig3}(b), and 
 the sum in Eq.~(\ref{Hpair}) runs over adjacent sites $\llangle klj\rrangle$ with $k<l<j$. The hopping amplitude for 
 singlet pairs is $\alpha J$ with $\alpha=1/2$. 
%
%
%%%%%%%%%%%%%%%%%%%%%%
%%%%%%%%%%%%%%%%%%%%
%

%
\begin{figure*}[t!]
\begin{center}
\includegraphics[width=1.4\columnwidth]{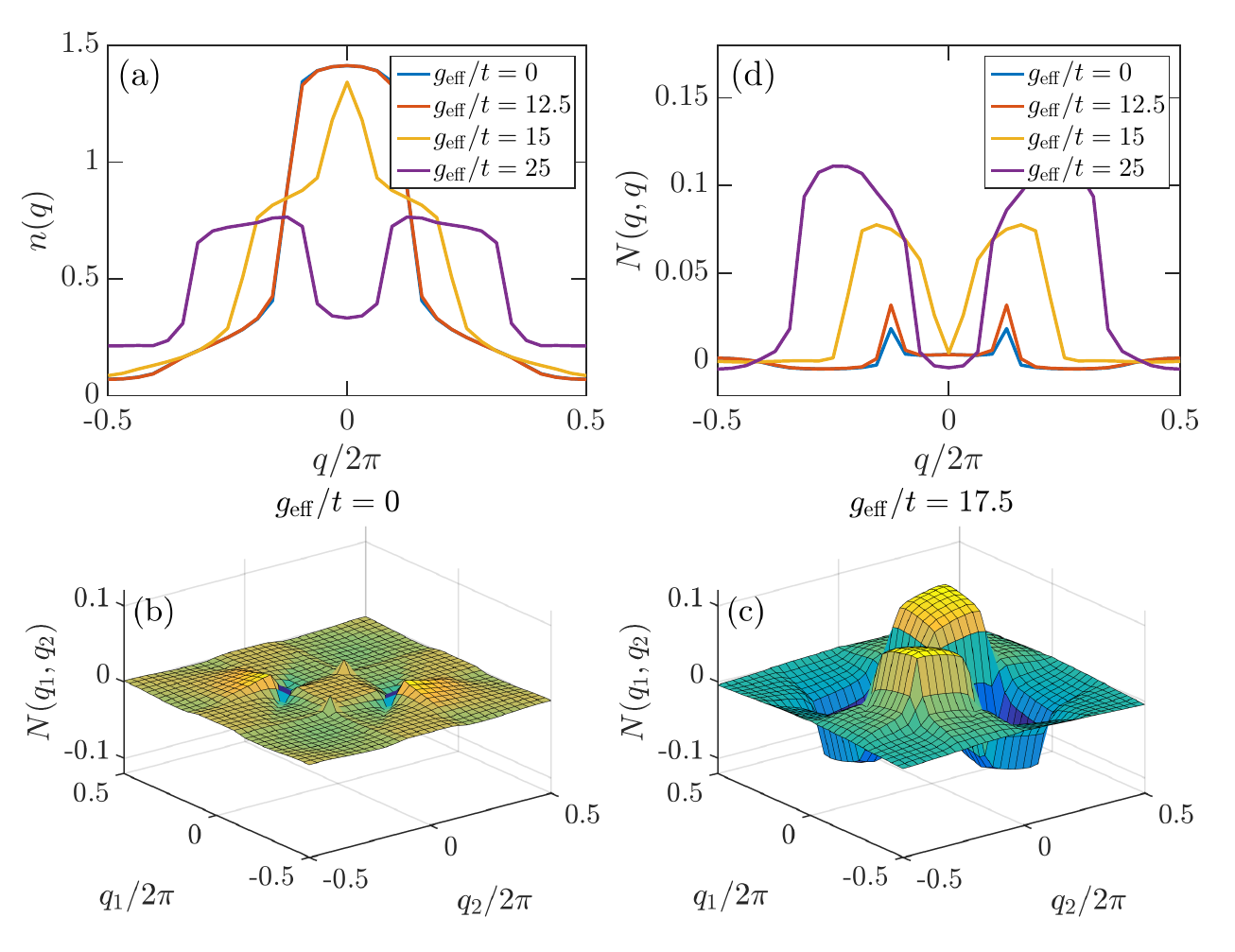}
\end{center}
\caption{\label{fig4}
(Color online) Illustration of the cavity-induced momentum correlations in the system. The model parameters are $U = \Omega = 20t$. The bond dimension used is $\chi = 400$, and the number of sites is $L=32$, and the system is at quarter-filling $N = 16$ electrons. 
(a) The momentum distribution of electrons $n(q)$ in the ground state at various values of the cavity coupling $g_{\rm eff}$.
(b) The momentum correlation $N(q_1,q_2)$ with a cavity coupling $g_{\rm eff}/t = 0$. 
(c) As in (b), but with a cavity coupling $g_{\rm eff}/t = 17.5$ after the transition point. 
(d) A diagonal slice $N(q,q)$ of the momentum correlations highlighting the qualitative difference before and after the transition point. 
}
\end{figure*} 
The presence of the cavity modifies the effective Hamiltonian in the manifold of zero excitations. The 
electron-photon interaction $\hat{V}$ couples $\hat{\mc{P}}_0$ to 
the $\hat{\mc{P}}_1$ and $\hat{\mc{P}}_2$ manifolds and  the effective Hamiltonian is [see Appendix~\ref{PertSeries}]
\begin{align}
  \hat{H}_{\text{gs}} = &  \hat{H}[tJ_c\alpha_c] + \hat{\mc{P}}_0\left(\hat{H}_{\text{shift}}  +\hat{H}_{\text{2-site}} + \hat{H}_{\text{long}}  \right) \hat{\mc{P}}_0\,. 
 \label{resg}
\end{align}
A comparison with the $tJ\alpha$ model shows that 
the cavity changes the parameters $J$ and $\alpha$ as shown in Fig.~\ref{fig3}(c).  
We find 
\begin{subequations}
 \begin{align}
 J_c  & = J\left(1+\mc{C}\right)\,, \\
\alpha_c &  = \alpha \frac{1-\mc{C}}{1+\mc{C}} \,,
 \end{align}
\end{subequations}
where 
\begin{align}
 \mc{C} & =  \frac{g^2}{t^2} \frac{U}{U+\Omega} \,. 
\end{align}
Since $\mc{C}>0$ for $g\not=0$,  the electron-photon interaction increases the magnetic interaction energy $J_c$, decreases $\alpha_c$  
and reduces the pair hopping amplitude 
\begin{align}
 \alpha_c J_c = \alpha J\left(1-\mc{C}\right)
\end{align}
for all parameters $\Omega,U\gg t, g$. Note that $J_c$ is the largest for the smallest $\Omega$ compatible with $\Omega\gg t, g$. 

In addition, the presence of the cavity results in three additional terms in Eq.~(\ref{resg}), 
\begin{subequations}
\label{add}
\begin{align}
& \hat{H}_{\text{shift}} =  - \frac{g^2}{\Omega}\sum\limits_j \left[\hat{n}_j (\mathds{1}-\hat{n}_{j+1}) 
+ \hat{n}_{j+1}(\mathds{1}-\hat{n}_j)\right] \,, \\
& \hat{H}_{\text{2-site}} =    \frac{g^2}{\Omega}\sum\limits_{j\sigma} \left[ (\mathds{1}-\hat{n}_{j}) c_{j-1,\sigma}^{\dagger}c_{j+1,\sigma} +  \text{h.c.} \right]\,, \\
& \hat{H}_{\text{long}} =  \frac{g^2}{\Omega}\sum\limits_{\begin{subarray}{c} k\not=l,l-1\\ \sigma, \nu \end{subarray}} 
\left[\hat{c}_{k,\sigma}^{\dagger}\hat{c}_{k+1,\sigma}\left(\hat{c}_{l,\nu}^{\dagger} \hat{c}_{l+1,\nu}
- \hat{c}_{l+1,\nu}^{\dagger} \hat{c}_{l,\nu}\right) +\text{h.c.}\right] ,
\end{align}
\end{subequations}
where   
$\hat{n}_j = \sum_{\sigma} \hat{n}_{j,\sigma}$. 
The physical processes induced by $\hat{H}_{\text{shift}}$, $\hat{H}_{\text{2-site}}$ and 
$\hat{H}_{\text{long}}$ are shown in Figs.~\ref{fig3}(d) and~(e). 
$\hat{H}_{\text{shift}}$ describes an energy shift of all singly-occupied sites  with an empty neighbouring site. 
This effective particle-hole binding energy results from  virtual transitions to the empty site  and are accompanied by the 
emission and re-absorption of virtual cavity photons.  $\hat{H}_{\text{2-site}}$ accounts for a next-nearest neighbour hopping process and 
$\hat{H}_{\text{long}}$ describes a long-range electron-electron interaction mediated by the cavity. 
These interactions are independent of the distance between electrons. 

We note that this technique of deriving an effective Hamiltonian is similarly applicable to classical driving fields analysed using Floquet theory \cite{mentink:15, eckardt:15, coulthard:17}. In a Floquet context, the zero-excitation sector is coupled to an infinite number of excited sectors $\hat{\mathcal{P}}_{m}$ with $-\infty \leq m \leq \infty$, representing absorption or emission of $m$ quanta from the driving field. However, the terms in the $+m$ and $-m$ sectors which contribute to $\hat{H}_{\rm shift}$, $\hat{H}_{\rm 2-site}$ and  $\hat{H}_{\rm long}$ exactly cancel in the case of a classical driving field. We conclude that as these terms do not appear for a classical driving field, they are quantum mechanical in origin. 

Finally, we note that we have carried out exact diagonalization calculations which confirm the 
accuracy of the effective Hamiltonian  $\hat{H}_{\text{gs}}$ in Eq.~(\ref{resg}) in the parameter regime $\Omega,U\gg t, g$, see Appendix~\ref{PertSeries} for details.
\subsection{Ground state properties \label{numerics}}
In order to determine the ground state of the combined electron-photon system we distinguish between electronic systems at half filling and below half filling. 

In the case of half filling the effective Hamiltonian in Eq.~(\ref{resg}) reduces to $\hat{H}_{\text{gs}}=\hat{\mc{P}}_0\hat{H}_{\text{mag}}[J_c]\hat{\mc{P}}_0$. 
This is the Hamiltonian of the $tJ\alpha$ model at half filling, i.e. an isotropic Heisenberg model with coupling $J_{c}$.
The ground state of $\hat{H}_{\text{gs}}$ is thus an antiferromagnetic state~\cite{essler:05} with increased exchange interaction $J_c > J$ due to the electron-photon coupling. 

To investigate the ground state at less than half filling, we perform finite-system DMRG calculations as implemented in the open source Tensor Network Theory (TNT) library~\cite{TNT}. The matrix product operator  corresponding to the Hamiltonian $H_{\text{gs}}$ is built using the finite automata technique~\cite{automata, paeckel:17}. We use a maximum matrix product state bond dimension of $\chi = 400$, resulting in a typical truncation error per bond of $\sim 10^{-4}$.

The most important term in the Hamiltonian determining the structure of the ground state is $\hat{H}_{\text{long}}$. Due to the infinite-range of the interaction, its total energy contribution dominates that of the strictly nearest-neighbour terms, scaling with $\sim L^2$ rather than $\sim L$. To account for this expected length-dependence of the coupling term, we define an effective cavity coupling 
\begin{equation}
g_{\rm eff} = 4 g \sqrt{L \ln{2}} \approx 3.33 g \sqrt{L}.
\end{equation}
To explore the effect of the cavity terms, we first compute the momentum distribution of electrons in the ground state,
\begin{equation}
n(q)  = \langle \sum_{\sigma} \hat{c}^{\dagger}_{q, \sigma} \hat{c}_{q, \sigma} \rangle,
\end{equation}
where $\hat{c}_{q, \sigma} = \sum_{j} {\rm e}^{-{\imag} j q} \hat{c}_{j, \sigma}/\sqrt{L}$. As shown in Fig.~\ref{fig4}(a), when $g_{\rm eff}/t = 0$ (i.e. the bare $tJ \alpha$ model), a distorted Fermi surface with the electrons centered about $q = 0$ is seen. When the cavity coupling is switched to sufficiently large values $g_{\rm eff}/t \gtrsim 15$, the Fermi surface splits into two smaller peaks at finite momenta. We find that when plotted as a function of the rescaled $g_{\rm eff}$, the transition occurs at the same value \emph{independent} of the system size $L$, as expected. 

Further information is revealed by looking at the momentum-space electron correlations,
\begin{equation}
N(q_1, q_2) = \langle c^{\dagger}_{q_1, \uparrow} c_{q_1, \uparrow} c^{\dagger}_{q_2, \downarrow} c_{q_2, \downarrow} \rangle
\end{equation}
which is shown in Fig.~\ref{fig4}(b) and Fig.~\ref{fig4}(c) for $g_{\rm eff}/t = 0$ and $g_{\rm eff}/t = 17.5$ respectively. We find that the cavity induced $\hat{H}_{\rm long}$ terms induce momentum-space pairing correlations such that pairs of “up" and “down" electrons always move in the same direction. In Fig.~\ref{fig4}(d) we show diagonal cuts $N(q,q)$ at various values of $g_{\rm eff}$, highlighting the induced correlations.  

In addition to $n(q)$ and $N(q_1, q_2)$ we calculated spin-spin 
correlations~\cite{coulthard:18} and find that they are approximately 
unchanged after the transition. This is because although $J_{c}$ and $\alpha_{c}$ are substantially modified by the presence of the cavity, they still lie below the required threshold to induce a magnetic phase transition \cite{coulthard:18}. The more important term, $\hat{H}_{\rm long}$ is spin agnostic, acting only on the charge degree of freedom. 

\section{Experimental realization \label{exp}}
Next  we discuss the experimental observation of the predicted effects. 
The  change in the magnetic interaction energy in the ground state at half filling
could, e.g., be  observed by measuring changes in the magnetic susceptibility $\Delta \chi \propto \Delta J\propto \eta^2$ of the material~\cite{tamura:02}. 
In order to predict the magnitude of this effect mediated by the exchange of virtual photons,  we consider $\text{ET-F}_2\text{TCNQ}$~\cite{hasegawa:97,hasegawa:00} which 
is a generic example of a one-dimensional Mott insulator where $U \gg t$. A cavity  mode with wavelength $\lambda_c\approx 1.8\mu\text{m}$ corresponding to 
the Mott gap $U\approx 0.7\,\text{eV}$ in $\text{ET-F}_2\text{TCNQ}$  results in 
$\eta \approx 3\times 10^{-5} \sqrt{\lambda_c^3/v}$.  This shows that significant coupling strengths require nanoplasmonic cavities~\cite{keller:17}  
where small values of $v/\lambda_c^3$ can be achieved. In order to change $J$ by $\approx 1\%$, we require $v/ \lambda_c^3\lesssim 10^{-7}$. 
Such small cavity volumes have been experimentally achieved recently~\cite{keller:17} for wavelengths in the THz regime. 
Similarly small volumes for higher frequencies are expected to be available in the near future, e.g., by using superconducting cavities~\cite{scalari:14,scalari:14b}. 

In order to observe the  cavity-induced pairing in momentum space at less than half filling we require 
$g_{\text{eff}}/t  \gtrsim 15$. 
If the material fills the mode volume of the cavity, $g_{\text{eff}}$ just depends on the volume per lattice site $v_{\text{uc}}$. 
Nanoplasmonic cavities are thus not required to observe the momentum-space pairing effect. 
In the case of $\text{ET-F}_2\text{TCNQ}$ we find 
 $g_{\text{eff}}/t \approx 6.4$. While this value 
 is too small by about a factor of two in order to 
 observe the new phase, larger values of 
 $g_{\text{eff}}/t$ are possible in materials with a smaller Mott gap or smaller unit cells. 
\section{Summary and discussion \label{conclusion}}
In summary, we have shown that second-order electron-photon interactions enhance superexchange 
interactions giving rise to the antiferromagnetic ground state 
of the one-dimensional Fermi-Hubbard model at half filling. 

Moreover, we have shown that at sufficiently large couplings, the cavity induces fermion-pairing in momentum space. In higher dimensions we speculate that this could lead to cavity-induced superconductivity~\cite{schlawin:18}.
Such effects do not emerge from a Floquet analysis of a classical light field, and so are genuinely quantum mechanical in nature. Similarly, substantial modification of the superexchange $J$ can only be achieved by classical light fields when they are extremely intense~\cite{mentink:15,coulthard:17,goerg:18}. Here this is achieved by strong coupling to an empty cavity.

Finally, we note that our results are directly applicable to higher-dimensional electronic systems where the electron-cavity interaction can be tuned via 
the relative orientation between the crystal and the cavity polarization vector, see  Appendix~\ref{elphot}. The rich physics ensuing from this anisotropic interaction is subject to future studies.
\begin{acknowledgments}
MK and DJ acknowledge financial support from the National Research Foundation
and the Ministry of Education, Singapore. DJ and FS acknowledge funding from the 
European Research Council under the European Union's Seventh Framework Programme 
(FP7/2007-2013)/ERC Grant Agreement no. 319286, Q-MAC. 
\end{acknowledgments}
%
%%%%%%%%%%%%%%%%%%%%%%%%%%%%%%%%%%%%%%%%
%
\appendix
\section{Electron-Photon interaction \label{elphot}}
The  Hubbard Hamiltonian  $\hat{H}_{\text{FH}}$ and the cavity Hamiltonian  $\hat{P}$ 
are defined in Eqs.~(\ref{HFH}) and~(\ref{hp}), respectively. Interactions between the two 
sub-systems
can be accounted for via the Peierls substitution~\cite{essler:05} $\hat{T}\rightarrow  \hat{T}_{\text{PS}}$, 
where  
\begin{align}
 \hat{T}_{\text{PS}} = &  -t \sum\limits_{ \langle j k \rangle \sigma} & 
 \left(  \hat{c}_{j,\sigma}^{\dagger}\hat{c}_{k,\sigma}
 e^{\imag  \frac{e}{\hbar} \int_{\mf{r}_j}^{\mf{r}_k}\mf{\hat{A}}(\mf{r}')\cdot\text{d}\mf{r}'} 
 +\text{h.c.} \right)\,, 
 \end{align}
$\mf{r}_i$ is the position vector of site $i$ and 
\begin{align}
 \mf{\hat{A}} = A_0 (a + a^{\dagger})\mf{u}
 \label{A}
\end{align}
is the quantized vector potential of the cavity field in Coulomb gauge. In Eq.~(\ref{A}) $\mf{u}$ is the mode function of the cavity field, 
 \begin{align}
 A_0 = \sqrt{\frac{\hbar}{2\varepsilon_0 \omega_c v}} \,,
  \label{avec}
 \end{align}
$\varepsilon_0$ is the vacuum permittivity and $v$ is the mode volume. 
In the following we neglect the position dependence of the mode function and assume that $\mf{u}=\mf{e}_c$, where 
$\mf{e}_c$ is the unit polarization vector of the cavity field. 
 Assuming that $\mf{e}_c$ is aligned with the direction of the atomic chain 
we obtain
 \begin{align}
 \hat{T}_{\text{PS}} =
 &  -t \sum\limits_{ \langle j k \rangle \sigma} & 
 \left(  \hat{c}_{j,\sigma}^{\dagger}\hat{c}_{k,\sigma}
 e^{\imag  \eta(\hat{a} +\hat{a}^{\dagger})} +\text{h.c.} \right)\,,
 \label{peierls}
 \end{align}
where the dimensionless parameter $\eta$ is defined in Eq.~(\ref{eta}) 
and $d = |\mf{r}_{j+1} - \mf{r}_j|$ is the 
lattice constant. The full Hamiltonian of the hybrid system comprising the electrons, 
photons and their interactions is thus 
\begin{align}
 \hat{H}_{\text{hybrid}} =   \hat{T}_{\text{PS}} + \hat{P} + \hat{D}\,.
 \label{Hhybrid}
 \end{align}
Expanding Eq.~(\ref{peierls}) up to first order in $\eta$ for $\eta\ll 1$  results in 
\begin{align}
  \hat{T}_{\text{PS}} \approx \hat{T} + \hat{V}\,,
\end{align}
where $\hat{V}$ is defined in Eq.~(\ref{Vint}). 
is the dimensionless current operator. 
For $\eta\ll 1$,  $\hat{H}_{\text{hybrid}}\approx \hat{H} = \hat{H}_{\text{FH}} + \hat{P}   + \hat{V} $  is thus well approximated by 
the system Hamiltonian $\hat{H}$ defined in  Eq.~(\ref{totv}). 
The generalization of the electron-photon interaction to higher-dimensional electronic systems is straightforward. 
In particular, for a three-dimensional 
crystal whose unit cell is described by the lattice vectors $\mf{d}_{\alpha}$ we obtain
\begin{align}
 \hat{V}_{\text{3D}} =  (\hat{a}+\hat{a}^{\dagger}) \frac{ e}{\sqrt{2\hbar\varepsilon_0 \omega_c v}}  
 \sum_{\alpha}t_{\alpha} (\mf{e}_c \cdot \mf{d}_{\alpha}) \hat{\mc{J}}_{\alpha}\,,
 \label{v3d}
\end{align}
where $t_{\alpha}$ and $\hat{\mc{J}}_{\alpha}$ are the hopping amplitude and the current operator  
in the direction of $\mf{d}_{\alpha}$, respectively. 
It follows  that the electron-cavity coupling depends on 
the relative orientation between the crystal and the cavity field. 
Note that we did not employ the rotating-wave approximation in Eqs.~(\ref{Vint}) and~(\ref{v3d}). As a matter of fact, the counter-rotating terms   give rise to 
the dominant contribution to the cavity-mediated effects in the ground state manifold.
\section{Evaluation of $H_{\text{gs}}$ \label{PertSeries}} 
%
%
%%%%%%%%%%%%%%%%%%%%%%
\begin{figure}[t!]
\begin{center}
\includegraphics[width=\columnwidth]{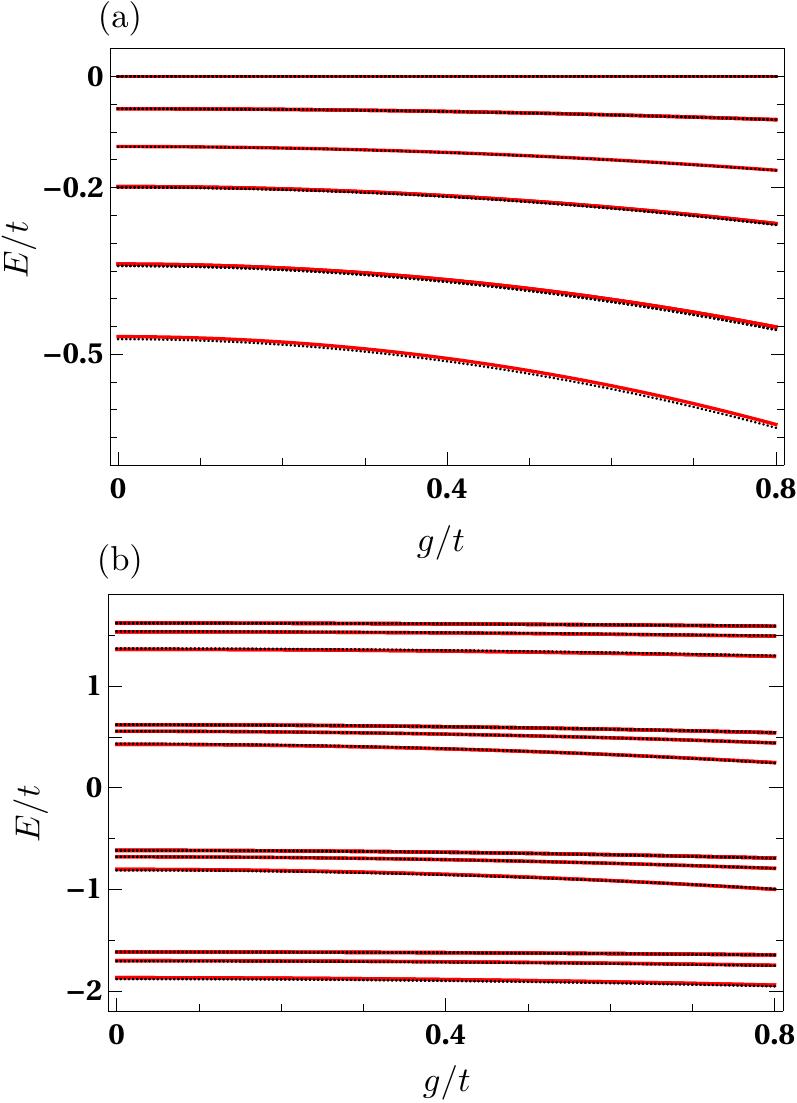} 
\end{center}
\caption{\label{fig5}
(Color online) Comparison of the eigenenergies $E$ of the system Hamiltonian $\hat{H}$ in Eq.~(\ref{totv}) in the ground state manifold and 
the effective Hamiltonian $\hat{H}_{\text{gs}}$ in Eq.~(\ref{resg}) for a system with $L=4$ sites as a function of the cavity coupling $g$. 
The eigenvalues corresponding to $\hat{H}$ ($\hat{H}_{\text{gs}}$) are shown by red solid (black dotted) lines. 
The exact diagonalization calculations take into account photon states $\ket{j_P}$ with $j \in \{0,1,2\}$. 
(a) Half filling with $N =4$ electrons and $U = 20 t$ and  $\Omega =18 t$. 
(b) Same as in (a) but for $N =3$ electrons. 
}
\end{figure} 
%%%%%%%%%%%%%%%%%%%%
% 
The first term in Eq.~(\ref{gs}) reduces to $\hat{\mc{P}}_0 \hat{T} \hat{\mc{P}}_0$ since $\hat{\mc{P}}_0 \hat{V} \hat{\mc{P}}_0 = 0$, i.e., 
the first-order contribution of the electron-photon coupling to $\hat{H}_{\text{gs}}$ vanishes. The sum in Eq.~(\ref{gs}) accounts for all 
second-order processes, and only the three terms with indices $(m=1,j=0)$,  $(m=2,j=1)$ and $(m=1,j=1)$ make a non-zero contribution to this sum. 
$(m=1,j=0)$  corresponds to the standard $tJ\alpha$ model, and $(m=2,j=1)$ and $(m=1,j=1)$ account for the modifications due to the cavity. 
In the following we discuss these three contributions in more detail:
\begin{itemize}
\item $(m=1,j=0)$: This term corresponds to processes where one doublon is created and subsequently annihilated. These processes 
only involve the hopping term  $\hat{T}$ and are independent of the electron-photon interaction $\hat{V}$, 
 \begin{align}
 & -\frac{1}{U} \hat{\mc{P}}_0 \hat{H}_1 \hat{\mc{P}}_1^{(0)} \hat{H}_1 \hat{\mc{P}}_0  = 
 -\frac{1}{U} \left[\hat{\mc{P}}_0^{D}  \hat{T} \hat{\mc{P}}_1^{D} \hat{T} \hat{\mc{P}}_0^{D}\right] \otimes \hat{\mc{P}}_0^{P}  \notag \\
 & =  \hat{\mc{P}}_0\left(\hat{H}_{\text{mag}}[J]  + \hat{H}_{\text{pair}}[\alpha J]   \right) \hat{\mc{P}}_0\,,
 \label{t10}
 \end{align}
where $\hat{H}_{\text{mag}}[J]$ and $\hat{H}_{\text{pair}}[\alpha J]$ are defined in Eqs.~(\ref{Hex}) and~(\ref{Hpair}), respectively, 
 $J=4  t^2/U$ and $\alpha = 1/2$. 
\item  $(m=2,j=1)$: This term accounts for the virtual creation and subsequent annihilation of one photon and one doublon, 
 \begin{align}
&  -\frac{1}{\Omega + U} \hat{\mc{P}}_0 \hat{H}_1 \hat{\mc{P}}_2^{(1)} \hat{H}_1 \hat{\mc{P}}_0 = 
 -\frac{g^2}{\Omega + U} \left[\hat{\mc{P}}_0^{D}  \hat{J} \hat{\mc{P}}_1^{D} \hat{J} \hat{\mc{P}}_0^{D}\right] \otimes \hat{\mc{P}}_0^{P} \notag \\
 &  =  \hat{\mc{P}}_0\left(\hat{H}_{\text{mag}}\left[\frac{4 g^2}{U+\Omega}\right]  + \hat{H}_{\text{pair}}\left[-\frac{2 g^2}{U+\Omega}\right]   \right) \hat{\mc{P}}_0\,.
  \label{t21}
 \end{align}
The terms in Eq.~(\ref{t21}) result in a re-normalization of the magnetic exchange energy and the pair hopping of the $tJ\alpha$ model. 
\item $(m=1,j=1)$:   This term describes processes where an electron 
 hops to a neighbouring empty site and a photon is emitted, followed by the re-absorption of the photon and a second hopping process. We find 
 \begin{align}
 &-\frac{1}{\Omega} \hat{\mc{P}}_0 \hat{H}_1 \hat{\mc{P}}_1^{(1)} \hat{H}_1 \hat{\mc{P}}_0 = 
- \frac{g^2}{\Omega} \left[\hat{\mc{P}}_0^{D} \hat{\mc{J}} \hat{\mc{P}}_0^{D}\hat{\mc{J}} \hat{\mc{P}}_0^{D}\right] \otimes \hat{\mc{P}}_0^{P} \notag \\
 & = \hat{\mc{P}}_0\left(\hat{H}_{\text{shift}}  +\hat{H}_{\text{2-site}} + \hat{H}_{\text{long}}  \right) \hat{\mc{P}}_0\,,
  \label{t11}
 \end{align}
where $\hat{H}_{\text{shift}}$, $\hat{H}_{\text{2-site}}$ and $\hat{H}_{\text{long}}$  are defined in Eq.~(\ref{add}).
Each process in Eq.~(\ref{t11}) involves two electron hops without creating a doublon. Depending on whether the two hopping processes 
go in the same or opposite direction, one obtains a particle-hole binding effect $(\hat{H}_{\text{shift}})$ or a next-nearest neighbour tunneling term ($\hat{H}_{\text{2-site}}$). 
The virtual photon can even be emitted and absorbed by two different electrons, which gives rise to the cavity-mediated long-range interaction $\hat{H}_{\text{long}}$.  
Note that all terms in Eq.~(\ref{t11}) are zero at half filling, since in this case electron hops without creating a doublon are impossible. 
\end{itemize}
Combining all terms in Eqs.{(\ref{t10}),~(\ref{t21}) and~(\ref{t11}) gives  $\hat{H}_{\text{gs}}$ in Eq.~(\ref{resg}).
In order to test the accuracy of this effective Hamiltonian, we compare its spectrum to the eigenvalues of the system Hamiltonian in Eq.~(\ref{totv}) via exact diagonalization. 
We find that the eigenvalues of the two Hamiltonians are in excellent agreement for sufficiently large values of $U$ and $\Omega$, and for a wide range of coupling strengths $g$. More specifically, 
the differences between the eigenvalues are of the order of $t^4/[\text{min}(U,\Omega)]^2$, 
which is the magnitude of the next higher-order terms in the perturbation series. We present two examples  of these calculations for a system with $L=4$ sites in  Fig.~\ref{fig5}, 
where Fig.~\ref{fig5}(a) and Fig.~\ref{fig5}(b) correspond to  half filling ($N=4$ electrons) and less than half filling ($N=3$ electrons), respectively. 
Note that we  chose a small system and an unrealistically large range of the cavity coupling parameter $g$ for illustration purposes.
Finally, we point out that some of the eigenvalues shown in Fig.~\ref{fig5} are degenerate. The total number of states in Figs.~\ref{fig5}(a) and (b) are 16 and 32, respectively. 
\end{document}